# Unprecedent fast winking of solar flares triggered by bursty magnetic reconnection

Ting Li[1,2,3], Xuchun Duan[2,3], Yijun Hou[2,3], Guillaume Aulanier[4,5,6], I.V. Zimovets[7], Jun Zhang[8], Juraj Lörinčík[9,10], Larisa Kashapova[11], Zhentong Li[12], Yining Zhang[2,3], Yulei Wang[13,14], Leping Li[2], Suli Ma[1], Jing Huang[2,3], Shuhong Yang[2,3] & Guiping Zhou[2,3]

## ABSTRACT

Flare ribbons form as a result of energy deposition associated with particles accelerated in low layers of the solar atmosphere. The fine-scale structures of flare ribbons, also called ribbon kernels, offer a potentially powerful diagnostic of the flare reconnection process, however to date the dynamic evolution of ribbon

---

[1]State Key Laboratory of Solar Activity and Space Weather, National Space Science Center, Chinese Academy of Sciences, Beijing 100190, China; liting@nssc.ac.cn

[2]CAS Key Laboratory of Solar Activity, National Astronomical Observatories, Chinese Academy of Sciences, Beijing 100101, China

[3]School of Astronomy and Space Science, University of Chinese Academy of Sciences, Beijing 100049, China

[4]Sorbonne Université, Observatoire de Paris-PSL, École Polytechnique, Institut Polytechnique de Paris, CNRS, Laboratoire de physique des plasmas (LPP), Paris, France

[5]Rosseland Center for Solar Physics, University of Oslo, Blindern, Norway

[6]Institute of Theoretical Astrophysics, University of Oslo, Blindern, Norway

[7]Space Research Institute of the Russian Academy of Sciences (IKI), 84/32 Profsoyuznaya Str, Moscow 117997, Russia

[8]School of Physics and Optoelectronics Engineering, Anhui University, Hefei 230601, China

[9]IBay Area Environmental Research Institute, NASA Research Park, Moffett Field, CA, USA

[10]Lockheed Martin Solar and Astrophysics Laboratory, Palo Alto, CA, USA

[11]Institute of Solar-Terrestrial Physics SB RAS, 126a Lermontov Str, Irkutsk 664033, Russia

[12]Key Laboratory of Dark Matter and Space Astronomy, Purple Mountain Observatory, Chinese Academy of Sciences, Nanjing 210023, China

[13]Institute of Science and Technology for Deep Space Exploration, Suzhou Campus, Nanjing University, Suzhou, 215163, People's Republic of China

[14]State Key Laboratory of Lunar and Planetary Sciences, Macau University of Science and Technology, Macau, People's Republic of China

kernels has not been fully characterized in statistical studies. Here, we checked the state-of-the-art observations (cadence $\leq$ 2.5 seconds) of solar flares in the ultraviolet from space by Interface Region Imaging Spectrograph (IRIS) over the past 12 years. Our results showed the first statistical study of multiple spatially-resolved flare kernel quasi-periodic pulsation events for 31 flares, with the period of 6-24 seconds. The ribbon kernels have a spatial scale of $480-1200$ km and some kernels exhibit unprecedent fast "winking" process, i.e., quasi-periodic pulsation-like flashing of individual kernels. The shortest heating time reaches about $2-3$ s, implying that the energy is deposited only in a small localized region within flare ribbons, persisting for only a few seconds. Meanwhile, some ribbon kernels were observed to slip along the ribbon at speeds of 20-1800 km s$^{-1}$. These observations strongly imply a joint picture for the dynamics and the bursty nature of ribbon kernels as being due to coupled effects of plasmoid formation and three-dimensional (3D) magnetic reconnection in the overlaying coronal current sheet. We suggest that the observed flare behaviors provide strong observational evidences of 3D bursty reconnection.

## 1. Introduction

Solar flares are one of the most direct consequences of magnetic reconnection in our Solar System, which release as much as $10^{32}$ ergs in a matter of minutes, causing enhanced emission across the entire electromagnetic spectrum. According to two-dimensional (2D) standard flare models (Lin & Forbes 2000; Shibata & Magara 2011), magnetic reconnection occurs at magnetic X-points below erupting coronal structures, followed by the formation of parallel flare ribbons along the polarity inversion line. Flare ribbons form as a result of energy deposition associated with particles accelerated in low layers of the solar atmosphere. High spatial resolution observations have showed that chromospheric flare ribbons are composed of numerous bright blob-like kernels with widths of about 100-500 km (Sobotka et al. 2016; Thoen Faber et al. 2025; Yadav et al. 2025). Recently, French et al. (2025a) referred to the ribbon kernel features observed in extreme ultraviolet (EUV) wavelength as bead-like structures with a key spatial separation of $1.7-1.9$ Mm. In addition to ribbon blob-like kernels, other fine-scale structures at flare ribbons such as the swirls and riblets have also been presented (Brannon et al. 2015; Parker & Longcope 2017; Pietrow et al. 2024; Singh et al. 2025).

Magnetic reconnection is intrinsically a three-dimensional (3D) process, which reveals observational features that can not be addressed by the conventional 2D model or its extensions (Schindler et al.1988; Pontin 2011; Li et al. 2021). Recent theoretical and 3D magnetohydrodynamic (MHD) simulation results showed that the formation of current layers, where reconnection occurs, is associated with quasi-separatrix layers (QSLs), regions characterized by strong gradients in magnetic connectivity while maintaining continuity (Demoulin et al. 1996; Titov et al. 2002; Kumar et al. 2021; Jiang et al. 2021; Pontin & Priest 2022; Dudík et al. 2025). Reconnection within the QSLs is of a slipping nature, which has also been referred to as slipping or slip-running reconnection (Aulanier et al. 2005, 2006; Janvier et al. 2013), depending on whether the slipping velocity of reconnecting field lines surpasses the coronal Alfvénic speed (ranging between a few hundred up to >1000 km $s^{-1}$). It has been found that flare ribbons correspond well with the photospheric traces of QSLs (Savcheva et al. 2015; Zhao et al. 2016) and ribbon kernels are observed to slip along the ribbons at sub-Alfvénic speed (Dudík et al. 2014; Li & Zhang 2015; Li et al. 2018). High ($\sim$ 2s) cadence observations of *Interface Region Imaging Spectrograph (IRIS)* (De Pontieu et al. 2014) recently revealed that flare ribbon kernels can move at super-Alfvénic speeds, a direct evidence of slip-running reconnection (Lörinčík et al. 2025; Zhang et al. 2025).

Magnetic reconnection essentially occurs at small scales, on the order of tens of meters (Litvinenko 1996). Previous studies have showed that large-scale current can endure the tearing mode instability and break into multiple magnetic islands (Ni et al. 2015; Ye et al. 2020; Wang et al. 2021; Wyper & Pontin 2021; Yan et al. 2022; Hou et al. 2024). The processes of tearing-mode instability and coalescence of magnetic islands occur repetitively, leading to a dynamical evolution of the current sheet known as impulsive bursty reconnection (Bhattacharjee et al. 1983; McAteer et al. 2007), with reconnection turning on and off rapidly or varying between fast and slow rates either quasi-periodically or randomly. Impulsive bursty or plasmoid-mediated reconnection is thought to result in quasi-periodic particle acceleration and flare radio pulsations (Kliem et al. 2000; Jelínek et al. 2017; Kumar et al. 2025). Total emission in solar and stellar flares often displays a clear oscillatory pattern, with typical periods ranging from sub-second timescales to several minutes, termed quasi-periodic pulsations (QPPs) (Nakariakov & Melnikov 2009; Huang et al. 2016; Tian et al. 2016; Yuan et al. 2019; Chen et al. 2019; Zimovets et al. 2021; Ning et al. 2022; Li et al. 2025). Such temporal intermittencies are of great interest because they are related to the energy release and particle acceleration processes (Lu et al. 2021; Corchado Albelo et al. 2024; Ashfield et al. 2026; Lörinčík et al. 2026). Despite being discovered for several decades, there still lacks the statistical study of multiple spatially-resolved flare kernel QPP events. Thus understanding the relation between QPPs and impulsive bursty reconnection is a major challenge since it involves a wide coupling of temporal and spatial scales.

In order to reveal the relations of QPPs pattern, impulsive bursty reconnection, and 3D

reconnection, we exploited the state-of-the-art observations of solar flare dynamics in the ultraviolet from space. By analyzing the high-cadence ($\leq$ 2.5 s) observations from the *IRIS* between 2013-2025, we find that ribbon kernels in 31 flare events show intensity pulsations with a period of 6-24 s. Some spatially resolved kernels exhibit unprecedent fast “winking” process, i.e., quasi-periodic pulsation-like flashing of individual kernels. Meanwhile, some ribbon kernels were observed to slip along the ribbon at speeds of 20-1800 km s$^{-1}$.

## 2. Observational Data and Analysis

In this work, we primarily focused on analysis of IRIS slit-jaw imager (SJI) 1330 and 1400 Å data[1]. The 1330 Å passband is dominated by emission from the C II 1335 Å and 1336 Å lines formed in the upper chromosphere with contributions from the transition region. The SJI 1400 Å channel is dominated by the Si IV 1402.77 Å line formed in the middle transition region. The pixel size of 1330 and 1400 Å images is roughly 0.33$''$, which is about 240 km at the surface of the Sun. After investigating the high-cadence observations between 2013-2025, a total of 31 flare events (including 28 C-class, 2 M-class and 1 unrecorded solar flares; see Table 1) are found to show the intensity pulsations and the simultaneous apparent slipping motions of ribbon kernels. It needs to be noted that some flares are not investigated for their complex morphology, evolution or saturation effect. The observed cadence for these 31 events ranges from 0.77 s to 2.43 s. For each event, the integrated intensity profiles within the entire ribbon region and 3 smaller ribbon sub-regions are calculated (the 3 distinct subregions are selected only if their intensity profiles exhibit at least 3 peaks, which ensures a more reliable determination of the period). The pulsation period for each intensity profile was calculated by using wavelet analysis. The mean period for these 4 intensity profiles (the entire ribbon region and 3 sub-regions) was thought to be the pulsation period for each event.

We also use multiple-wavelength observations by the *Atmospheric Imaging Assembly (AIA)* (Lemen et al. 2012) on board the *Solar Dynamics Observatory (SDO)* spacecraft (Pesnell et al. 2012). The AIA takes full-disk images in 10 (E)UV channels with a pixel size of 0.6$''$ and a cadence of 12/24 s. We also used microwave observations from the *Siberian Radioheliograph (SRH)* (Altyntsev et al. 2020) and the *SOLARSPEL spectropolarimeter (SolSpel)* in the 3-24 GHz range as an indicator of accelerated particle presence. The combination of the spectral data from SRH and SolSpel provides a spectrum of microwave flare emissions. The Hard X-ray (HXR) emission detected by the *Hard X-ray Imager (HXI)* (Su et al. 2024) aboard the *Advanced Space-based Solar Observatory (ASO-S)* (Gan et al. 2023) carries information about energetic electrons accelerated by magnetic reconnection. We reconstructed the HXR image by using the CLEAN algorithm, utilizing data from detectors D29-D91 within the 13-20 keV range.

[1]https://iris.lmsal.com/search/

## 3. Observational Results

Among the 31 flare events that show the intensity pulsations and the simultaneous apparent slipping motions of ribbon kernels, two of them (Event 1 on 2024 August 25 and Event 2 on 2024 September 23) are selected as examples to analyze in detail.

### 3.1. Overview of Event 1 on 2024 August 25

One flare example under study originated in the National Oceanic and Atmospheric Administration (NOAA) active region (AR) 13796 on 2024 August 25. It was observed by the *IRIS* satellite at a cadence of about 0.87 s. The flare was a C3.0-class one recorded by the *GOES* and it peaked at about 09:50 UT (Figure 1(a)). Both peaks of the temporal profiles, at 6 GHz from the *SRH* and 1330 Å, integrated intensity within the ribbon region, occurred at about 09:48 UT. During the impulsive phase of the flare, the 1330 Å integrated intensity temporal profile showed evident intensity pulsations with a period of about 7.2 s (Figure 1(b)). The combined microwave spectrum from *SRH* and *SolSpel* in the 3-24 GHz range is characteristic of gyrosynchrotron emissions from non-thermal electrons (Figure 1(c)). *SDO* 171 Å and *IRIS* high-resolution 1330 Å images showed that the ribbon was composed of numerous bright kernels, suggesting that flare ribbons are "patchy" in nature (Figures 1 (d)-(e); French et al. 2021; Yadav et al. 2025). The sizes of these kernels range from 480 to 1200 km (2−5 pixels), and they are spaced roughly $\sim$ 1600 km apart at the onset of the flare.

### 3.2. "Winking" ribbon kernels of Event 1

The ribbon kernels exhibit the evident intensity oscillation during the impulsive phase of the flare and the detailed oscillation processes of two kernels are shown in Figure 2. At 09:47:30 UT, kernels "K1" and "K2" had the average intensity of 74 and 128 DN pixel$^{-1}$ within the rectangle regions (Figure 2(a1)). About 2.6 seconds later, the intensity of kernels "K1" and "K2" decreased by about 70% (Figure 2(a2)). Later, their intensity increased again by about 300% at 09:47:37 UT (Figure 2(a3)). Kernels "K1" and "K2" underwent the "bright-dim-bright-dim" intensity oscillation between 09:47:30 UT and 09:47:45 UT, with an average period of about 6 s (Figures 2(a1)-(a6)).

In order to highlight the "winking" process, the integrated intensity evolution within the regions of kernels "K1" and "K2" was calculated and displayed in Figure 2(b). The time profiles for kernels "K1" and "K2" both exhibited quasi-periodic variation. Their peak intensity increased by 100%−300% compared to the lowest integrated intensity and the two curves showed the overall ascending trend as the flare developed. To quantify the period of intensity oscillation, the time profiles of detrended intensity for kernels "K1" and "K2" and their Morlet spectrum are shown in Figures 2(c)-(d). The oscillation periods for "K1" and "K2" were about 6.1 s and 5.1 s, respectively.

### 3.3. Fast slippage of ribbon kernels of Event 1

Besides the observations of intensity oscillation, we also find that some ribbon kernels exhibit apparent slipping motions along the ribbon at some time. Two episodes of the apparent slippage were shown in Figures 3(a)-(d). Kernel "K3" was initially very bright at 09:47:09 UT, and then it started to slip towards the east and meanwhile its intensity decreased. The intensity in the neighboring locations "K4" and "K5" later increased in sequence. From kernel "K3" to kernel "K5", the overall slipping distance was about 2.8 Mm and the slippage speeds were 700±280 and 1800±550 km $s^{-1}$ (Figure 3(b)). The successive brightening of adjacent different locations can be attributed to the apparent slipping motion of the kernel along the ribbon. Different from the slipping direction of "K3"−"K5", kernel "K6" exhibited an apparent slipping motion towards the west, which caused the successive brightening of neighboring kernels "K7" and "K8" (Figure 3(c)). The slipping distance from kernel "K6" to "K8" was about 2.9 Mm and the slipping speeds were calculated to be 1700±550 and 1500±550 km $s^{-1}$ (Figure 3(d)). The stack plot along the ribbon exhibits multiple nearly vertical stripes, indicating the "winking" process of ribbon kernels (Figures 3(e)-(f)). The fitting of several inclined stripes shows the slipping speed of 1100-1500 km $s^{-1}$. It needs to be noted that determining if nearly vertical stripes are truly vertical (i.e., stationary intensity pulsations of kernels) or slightly inclined (i.e., very fast slippage) is difficult only based on the stack plot. However, by combining the ribbon evolution, we suggest that at least some kernels exhibit the fast apparent slippage. The slippage of ribbon kernels surpasses the coronal Alfvénic speed, and the super-Alfvénic slippage implies the occurrence of slip-running reconnection in the conditions when the QSLs are thin and neighboring field lines there have strong gradients of connectivity (Aulanier et al. 2006; Janvier et al. 2013).

### 3.4. Event 2 on 2024 September 23: “winking” and slipping process

Another flare event on 2024 September 23 was selected as an example for this event shows simultaneous super-Alfvénic and sub-Alfvénic slippage of ribbon kernels. The flare was a C4.5-class one recorded by the *GOES*, which originated in AR 13825 near the west limb. It started at 09:59 UT, peaked at 10:06 UT and ended at 10:10 UT (Figure 4(a)). The HXR emission from *HXI* aboard the *ASO-S* shows a significant enhancement during the flare impulsive phase at 10-20 keV HXR channel. The reconstructed HXR sources at 13-20 keV observed by *HXI* are co-spatial with the two flare ribbons observed by *AIA* at 1600 Å (Figure 4(b)). High-temperature flare loops connecting the two flare ribbons can be clearly discerned from *AIA* 131 Å observations (Figure 4(c)). IRIS observed the event at an cadence of $\sim$ 2.4 s. The 1330 Å integrated intensity within the ribbon region shows an oscillating pattern during the impulsive and decay phases of the flare (Figures 4(a) and 5(a)-(b)).

The ribbon kernels are observed to propagate along the ribbon and their slipping speed varies significantly at different positions (Figure 5). For the eastern ribbon, the slippage is towards the south and the slipping speed is about 20 km s$^{-1}$ at the north part, while it reaches about 300-800 km s$^{-1}$ at the south part (Figure 5(d)). The slipping distance is about 10 Mm, much larger than Event 1. The stack plot along the western ribbon shows multiple moving intensity features, each corresponding to the apparent slippage of ribbon kernels. These kernels exhibited fast motion during the initial stage, with an apparent velocity of about 260-1400 km s$^{-1}$. Subsequently, the slipping motion decelerated and the velocity gradually decreased to 70-100 km s$^{-1}$ (Figure 5(c)). Both the stack plot and the light curves along two horizontal lines exhibit the clear quasi-periodic pattern (Figures 5(c), 5(e)-(f)). For a fixed ribbon position, we can see that the kernels show the clear “bright-dim-bright-dim” pattern, with a period of about 17 s (Figure 6). Usually after the emission intensity of the kernel reached its peak, the kernel then slipped along the ribbon and the intensity at the previous bright kernel location decreased accordingly. Among the 31 flare events, other 3 events (2025 February 05, 2025 April 05 and 2024 November 23) are shown in Appendix. The histogram of detected periods of QPPs for 31 flare events is presented in Figure 7. The period of QPPs in flare ribbons observed in far UV band is within the range of 6−24 s and the mean period is 14.8 s (vertical green line).

## 4. Summary and Discussion

Thanks to the high temporal ($\leq$ 2.5 s) and spatial ($0.33''$ pixel$^{-1}$) resolutions of *IRIS*, we have observed the prevalent existence of short-period intensity pulsations with the period of about 6-24 s and slipping motion of flare ribbon kernels at speeds of 20-1800 km s$^{-1}$.

Some spatially resolved kernels (with spatial size of 480−1200 km) exhibit unprecedent fast "winking" process. We suggest that the "winking" process of ribbon kernels provides the evidence for bursty magnetic reconnection (Priest 1985; Tajima et al. 1987; McAteer et al. 2007), with the reconnection process being significantly influenced by the dynamics of the plasmoid instability, which involves the growth of plasmoids, their ejection from the current layer, plasmoid coalescence, and the accompanying "secondary tearing" of current sheets (Kliem et al. 2000; Jelínek et al. 2017; Kumar et al. 2025).

These high-cadence observations strongly imply a dual nature of the reconnection process in the overlaying coronal current sheet, involving the plasmoid-mediated and slipping/slip-running reconnection. A simplified schematic picture of the coupled plasmoid-mediated and slipping/slip-running reconnection is provided in Figure 8. From time "t0" to "t6", the magnetic field lines (colored lines) undergo a continuous series of slipping/slip-running reconnections and gradually change in connectivity. Their footpoints correspond to the observed ribbon kernels and exhibit the apparent slipping motion towards the south (pink arrow). The final state of the system is that the pre-reconnection field line "1" (long blue line) changes its connectivity into that of the pre-reconnection field line "2" (long red line). For a fixed ribbon position, the "winking" phenomenon implies the bursty and dynamic nature due to the formation and coalescence of plasmoids (pink and purple twisted field lines) at the current sheet. According to the recent MHD simulations (Dahlin et al. 2025), the formation and propagation of the plasmoid toward the flare arcade manifests as the "filling in" of the spirals, with a morphology consistent with observed fine ribbon structure. Thus we suggest that the "winking" ribbon kernels reflect the generation, propagation,and annihilation of the plasmoids.

The period of 6−24 s is consistent with the peak QPP distribution period of 20 s in the statistical study of GOES X-ray light curves (Hayes et al. 2020). Radio sources at flare loops and current sheet display the similar QPP pattern with a period of about 5−60 s, which is suggested to arise from the modulation of magnetic islands within the current sheet (Luo et al. 2022; Kou et al. 2022). Similar QPP period has also been reported in HXR emission and the $\gamma$-ray band related to nonthermal ions (Li et al. 2020; Collier et al. 2024), which is believed that spontaneous quasi-periodic energy release is probably the driver of the observed pulsations. Recently, French et al. (2025b) and Ashfield et al. (2026) found that Doppler velocities in flare ribbons exhibit QPP-like oscillations with periods of $\sim$ 50 s and 32 s respectively, which are slightly larger than that in our statistical results.

Although QPPs of similar periods have been reported in many previous studies, the statistical study of multiple spatially-resolved flare kernel QPP events has never been carried out in previous observations with lower temporal and spatial resolution. Our results present the first statistical study of multiple spatially-resolved flare kernel QPP events for 31 flares. The formation, motion, and coalescence of magnetic islands with each other are found to generate fast magnetoacoustic waves (Yang et al. 2015; Mondal et al. 2024). Recent theoretical and numerical simulation results showed that the bursty reconnection can drive natural modes of oscillation of the current sheet (Artemyev & Zimovets 2012; Jelínek et al. 2017; Mondal et al. 2025), with the periods determined by the Alfvénic travel time within the plasmoids. We suggest that the excited fast magnetoacoustic waves or the oscillations of the current sheet during the process of bursty reconnection are able to modulate reconnection rate and the precipitation of nonthermal particles at the footpoints of the reconnecting flare loops, resulting in the observed intensity oscillation of ribbon kernels.

The slippage of ribbon kernels is only an apparent motion and does not correspond to real bulk motions, but to the rearrangement of the global field lines as a result of reconnection within QSLs (Aulanier et al. 2006; Janvier et al. 2013). The slipping speed varies between different events, and even at the same event, there is a speed discrepancy for different parts of the ribbon or at different stages of the flare evolution (Figure 5). The slipping velocities of ribbon kernels were found to be related with the strength and distribution of the magnetic field (Lörinčík et al. 2019). The slipping distance is estimated to be 3-10 Mm, only a portion of the ribbon, not the entire flare ribbon. he ribbon kernels observed in far UV have a spatial scale of about 480−1200 km and the shortest heating time of 2−3 s (Figure 2), similar to that observed in EUV band (French et al. 2025a; Collier et al. 2026). This suggests that the energy was deposited only in a small localized region within flare ribbons, persisting for only a few seconds.

Our study presents both the localized "winking" kernels and their apparent slippage, which further requires a combined interpretation involving localized emission from energetic particles released from a plasmoid, coupled with a continuous change in connectivity due to sub-/super-Alfvénic field line slippage, i.e. slipping/slip-running reconnection. The observed kernel dynamics and oscillatory intensity evolution indicate that the dynamics of ribbon kernels is the response to 3D bursty reconnection in the coronal current sheet. Our observational results provide unprecedented evidences of footpoint traces of coupled plasmoid-mediated and slipping/slip-running reconnection in the corona during a solar flare. Thus they not only provide a coupling between the energy release and QPP patterns during solar flares, but also have important implications for understanding the physical mechanisms underpinning three-dimensional and bursty reconnection that can occur in a variety of situations, from astrophysical environments, laboratory experiments, and fusion devices.

This work is supported by the B-type Strategic Priority Program of the Chinese Academy of Sciences (grant No. XDB0560000), the National Natural Science Foundations of China

(12273060, 12333010, 12473057, 12573056 and 12533010), the Youth Innovation Promotion Association of CAS (2023063), the National Key R&D Program of China 2022YFF0503002, Beijing Natural Science Foundation (1252034), the Specialized Research Fund for State Key Laboratory of Solar Activity and Space Weather and the Prominent Postdoctoral Project of Jiangsu Province (2023ZB304). G.A. acknowledges the Appel à Proposition de Recherche of CNES/SHM and by the Action Thématique Soleil-Terre (ATST) of CNRS/INSU PN Astro, also funded by CNES, CEA, and ONERA. I.Z. is supported by the subsidy PLASMA. L.K. acknowledges the Ministry of Science and Higher Education of the Russian Federation and PIFI Group grant No. 2025PG0008. This research was supported by the International Space Science Institute (ISSI) in Bern, through ISSI International Team project (ISSI Team project #25-658). IRIS is a NASA small explorer mission developed and operated by LMSAL with mission operations executed at NASA Ames Research Center and major contributions to downlink communications funded by ESA and the Norwegian Space Agency. SDO data were obtained courtesy of NASA/SDO and the AIA and HMI science teams. We thank the SRH, SolSpel and ASO-S teams for providing observational data.

# APPENDIX

## A. Event 3 on 2025 February 05

An M1.1-class two-ribbon flare occurring in AR 13981 on 2025 February 05 was detected by *IRIS* SJI 1330 Å high-cadence mode with a cadence of about 0.87 s (Figure 9). Initially, the east part of the southern ribbon was illuminated before 03:14 UT. Then it started to slip towards the west at the speed of about 1300 km s$^{-1}$ (Figure 9(c)). This slippage episode can be discerned in 5 frames of 1330 Å images and the slippage distance was about 6 Mm. The slipping motion appeared intermittently and there exist multiple stripes in the stack plot along the southern ribbon. The integrated intensity profiles of 1330 Å at two fixed regions show distinct pulsation characteristics with a period of about 9 s (Figure 9(d)).

## B. Event 4 on 2025 April 05

Among these 31 flares, only two flares have circular ribbons and most of them are two-ribbon flares. In a circular-ribbon flare, null-point magnetic reconnection takes place, and accelerated-particles move from the null point along both the separatrix surface and the spine field into the lower atmosphere. On 2025 April 05, *IRIS* SJI 1330 Å observed a circular M1.0-class flare in AR 14048 with a cadence of about 2.27 s (Figure 10). The flare started at 19:54 UT, peaked at 20:05 UT and ended at 20:12 UT recorded by GOES. The flare is composed of a circular ribbon and an inner hook-shaped ribbon. As seen from the stack plot along the circular ribbon (Figure 10(c)), it shows multiple nearly-vertical stripes, with each stripe corresponding to one fast slipping episode. By fitting one slightly inclined stripe, the speed is about 1500 km $s^{-1}$. The integrated intensity profiles of 1330 Å at two fixed ribbon regions show an evident QPP pattern with a period of about 19 s (Figure 10(d)).

## C. Event 5 on 2024 November 23

On 2024 November 23, *IRIS* SJI 1330 Å observed a C3.8-class flare in AR 13901 with a cadence of about 0.97 s (Figure 11). The flare ribbon is composed of sawtooth fine structures, which is different from the dot-like kernels in Event 1. Starting from 21:44 UT, the ribbon substructures were observed to slip along the ribbon towards the northeast. The stack plot along the ribbon shows multiple moving intensity features, with each strip representing the apparent slippage of ribbon substructures. These stripes are not smooth probably for the sawtooth fine structures. This slippage episode can be discerned in more than 7 frames of 1330 Å images and the slippage distance was over 7 Mm. The speed of one slipping episode calculated by fitting is about 1100 km $s^{-1}$. For two fixed ribbon positions, 1330 Å intensity profiles exhibited short-period pulsations (approximately 11 s) (Figure 11(d)).

## REFERENCES

Altyntsev, A., Lesovoi, S., Globa, M., et al. 2020, Solar-Terrestrial Physics, 6, 2, 30.

Artemyev, A. & Zimovets, I. 2012, Sol. Phys., 277, 2, 283.

Ashfield, W., Polito, V., Lörinčík, J., et al. 2026, Nature Astronomy, 10, 54.

Aulanier, G., Pariat, E., & Démoulin, P. 2005, A&A, 444, 3, 961.

Aulanier, G., Pariat, E., Démoulin, P., et al. 2006, Sol. Phys., 238, 2, 347.

Bhattacharjee, A., Brunel, F., & Tajima, T. 1983, Physics of Fluids, 26, 11, 3332.

Brannon, S. R., Longcope, D. W., & Qiu, J. 2015, ApJ, 810, 1, 4.

Chen, X., Yan, Y., Tan, B., et al. 2019, ApJ, 878, 2, 78.

Collier, H., Hayes, L. A., Yu, S., et al. 2024, A&A, 684, A215.

Collier, H., Krucker, S., Hayes, L. A., et al. 2026, arXiv:2603.12178.

Corchado Albelo, M. F., Kazachenko, M. D., & Lynch, B. J. 2024, ApJ, 965, 1, 16.

Dahlin, J. T., Antiochos, S. K., Wyper, P. F., et al. 2025, ApJ, 993, 1, 31.

Demoulin, P., Henoux, J. C., Priest, E. R., et al. 1996, A&A, 308, 643.

De Pontieu, B., Title, A. M., Lemen, J. R., et al. 2014, Sol. Phys., 289, 7, 2733.

Dudík, J., Janvier, M., Aulanier, G., et al. 2014, ApJ, 784, 2, 144.

Dudík, J., Aulanier, G., Lörinčík, J., et al. 2025, Sol. Phys., 300, 10, 139.

French, R. J., Kazachenko, M. D., Berghmans, D., et al. 2025a, ApJ, 995, 2, L54.

French, R. J., Ashfield, W. H., Tamburri, C. A., et al. 2025b, ApJ, 995, 2, 182.

French, R. J., Matthews, S. A., Jonathan Rae, I., et al. 2021, ApJ, 922, 2, 117.

Gan, W., Zhu, C., Deng, Y., et al. 2023, Sol. Phys., 298, 5, 68.

Hayes, L. A., Inglis, A. R., Christe, S., et al. 2020, ApJ, 895, 1, 50.

Hou, Z., Tian, H., Madjarska, M. S., et al. 2024, A&A, 687, A190.

Huang, J., Kontar, E. P., Nakariakov, V. M., et al. 2016, ApJ, 831, 2, 119.

Janvier, M., Aulanier, G., Pariat, E., et al. 2013, A&A, 555, A77.

Jelínek, P., Karlický, M., Van Doorsselaere, T., et al. 2017, ApJ, 847, 2, 98.

Jiang, C., Chen, J., Duan, A., et al. 2021, Frontiers in Physics, 9, 575.

Kliem, B., Karlický, M., & Benz, A. O. 2000, A&A, 360, 715.

Kou, Y., Cheng, X., Wang, Y., et al. 2022, Nature Communications, 13, 7680.

Kumar, P., Karpen, J. T., & Dahlin, J. T. 2025, ApJ, 980, 2, 158.

Kumar, S., Nayak, S. S., Prasad, A., et al. 2021, Sol. Phys., 296, 1, 26.

Lemen, J. R., Title, A. M., Akin, D. J., et al. 2012, Sol. Phys., 275, 1-2, 17.

Li, D., Kolotkov, D. Y., Nakariakov, V. M., et al. 2020, ApJ, 888, 2, 53.

Li, D., Yuan, D., Yan, J., et al. 2025, Journal of Geophysical Research (Space Physics), 130, 4, e2025JA033772.

Li, T., Hou, Y., Yang, S., et al. 2018, ApJ, 869, 2, 172.

Li, T., Priest, E., & Guo, R. 2021, Proceedings of the Royal Society of London Series A, 477, 2249, 20200949.

Li, T. & Zhang, J. 2015, ApJ, 804, 1, L8.

Lin, J. & Forbes, T. G. 2000, J. Geophys. Res., 105, A2, 2375

Litvinenko, Y. E. 1996, ApJ, 462, 997.

Lörinčík, J., Aulanier, G., Dudík, J., et al. 2019, ApJ, 881, 1, 68.

Lörinčík, J., Dudík, J., Sainz Dalda, A., et al. 2025, Nature Astronomy, 9, 45.

Lörinčík, J., Collier, H., Polito, V., et al. 2026, arXiv:2604.17330.

Lu, L., Li, D., Ning, Z., et al. 2021, Sol. Phys., 296, 8, 130.

Luo, Y., Chen, B., Yu, S., et al. 2022, ApJ, 940, 2, 137.

McAteer, R. T. J., Young, C. A., Ireland, J., et al. 2007, ApJ, 662, 1, 691.

Mondal, S., Srivastava, A. K., Pontin, D. I., et al. 2024, ApJ, 977, 2, 235.

Mondal, S., Srivastava, A. K., Pontin, D. I., et al. 2025, ApJ, 989, 2, 222.

Nakariakov, V. M. & Melnikov, V. F. 2009, Space Sci. Rev., 149, 1-4, 119.

Ni, L., Kliem, B., Lin, J., et al. 2015, ApJ, 799, 1, 79.

Ning, Z., Wang, Y., Hong, Z., et al. 2022, Sol. Phys., 297, 1, 2.

Parker, J. & Longcope, D. 2017, ApJ, 847, 1, 30.

Pesnell, W. D., Thompson, B. J., & Chamberlin, P. C. 2012, Sol. Phys., 275, 1-2, 3.

Pietrow, A. G. M., Druett, M. K., & Singh, V. 2024, A&A, 685, A137.

Pontin, D. I. 2011, Advances in Space Research, 47, 9, 1508.

Pontin, D. I. & Priest, E. R. 2022, Living Reviews in Solar Physics, 19, 1, 1.

Priest, E. R. 1985, Reports on Progress in Physics, 48, 7, 955.

Savcheva, A., Pariat, E., McKillop, S., et al. 2015, ApJ, 810, 2, 96.

Schindler, K., Hesse, M., & Birn, J. 1988, J. Geophys. Res., 93, A6, 5547.

Shibata, K. & Magara, T. 2011, Living Reviews in Solar Physics, 8, 1, 6.

Singh, V., Scullion, E., Botha, G. J. J., et al. 2025, arXiv:2507.01169.

Sobotka, M., Dudík, J., Denker, C., et al. 2016, A&A, 596, A1.

Su, Y., Zhang, Z., Chen, W., et al. 2024, Sol. Phys., 299, 10, 153.

Tajima, T., Sakai, J., Nakajima, H., et al. 1987, ApJ, 321, 1031.

Thoen Faber, J., Joshi, R., Rouppe van der Voort, L., et al. 2025, A&A, 693, A8.

Tian, H., Young, P. R., Reeves, K. K., et al. 2016, ApJ, 823, 1, L16.

Titov, V. S., Hornig, G., & Démoulin, P. 2002, Journal of Geophysical Research (Space Physics), 107, A8, 1164.

Wang, Y., Cheng, X., Ding, M., et al. 2021, ApJ, 923, 2, 227.

Wyper, P. F. & Pontin, D. I. 2021, ApJ, 920, 2, 102.

Yadav, R., Kazachenko, M. D., Cauzzi, G., et al. 2025, ApJ, 989, 2, 183.

Yan, X., Xue, Z., Jiang, C., et al. 2022, Nature Communications, 13, 640.

Yang, L., Zhang, L., He, J., et al. 2015, ApJ, 800, 2, 111.

Ye, J., Cai, Q., Shen, C., et al. 2020, ApJ, 897, 1, 64.

Yuan, D., Feng, S., Li, D., et al. 2019, ApJ, 886, 2, L25.

Zhang, Y., Li, T., Hou, Y., et al. 2025, ApJ, 982, 1, L9.

Zhao, J., Gilchrist, S. A., Aulanier, G., et al. 2016, ApJ, 823, 1, 62.

Zimovets, I. V., McLaughlin, J. A., Srivastava, A. K., et al. 2021, Space Sci. Rev., 217, 5, 66.

---

This preprint was prepared with the AAS LaTeX macros v5.2.

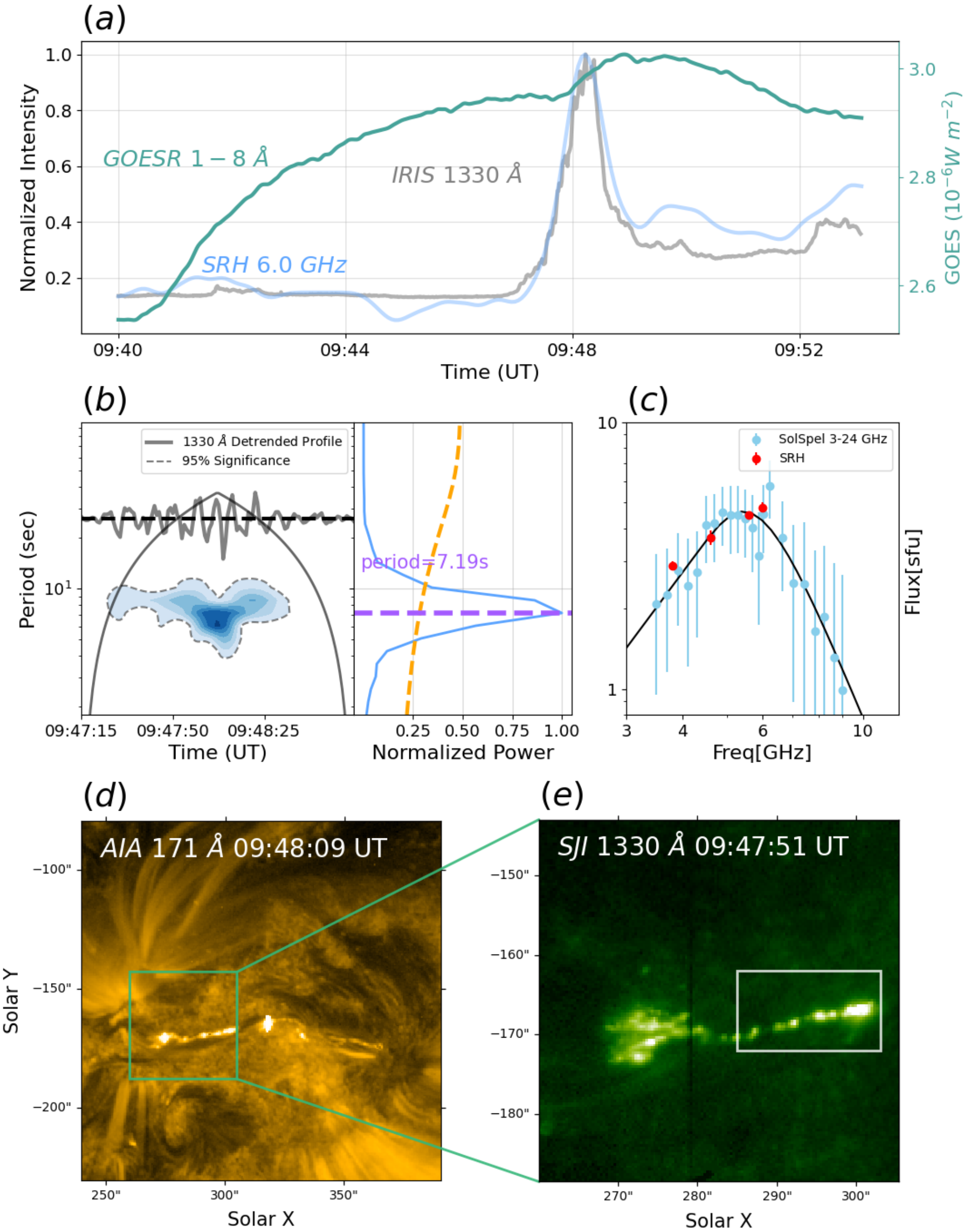

Fig. 1.— Context observations of Event 1 on 2024 August 25. (a) Time profiles of the flare in SRH microwave emission at 6 GHz, the soft X-ray flux in the 1−8 Å channel of the GOES satellite and the IRIS SJI 1330 Å integrated intensity within FOV of panel (e). (b) The wavelet analysis of 1330 Å integrated intensity, including the wavelet power map of the detrended intensity and corresponding global power. The orange dashed line represents the 95% confidence level, and the purple dashed line denotes the period over the 95% confidence level with the maximal power. (c) The combined microwave spectrum obtained at 09:48 UT. Red and blue circles respectively show the flux values derived from SRH and SolSpel data. The black line presents the results of fitting the observed microwave spectrum. (d) AIA 171 Å image showing the flare ribbons. (e) *IRIS* SJI 1330 Å image showing the east flare ribbon. The white rectangle in panel (e) is the FOV of SJI 1330 Å images in Figures 2-3. An animation showing the evolution of the flare ribbon from 09:46 to 09:50 UT is available. The real-time duration of the animation is 24 s.

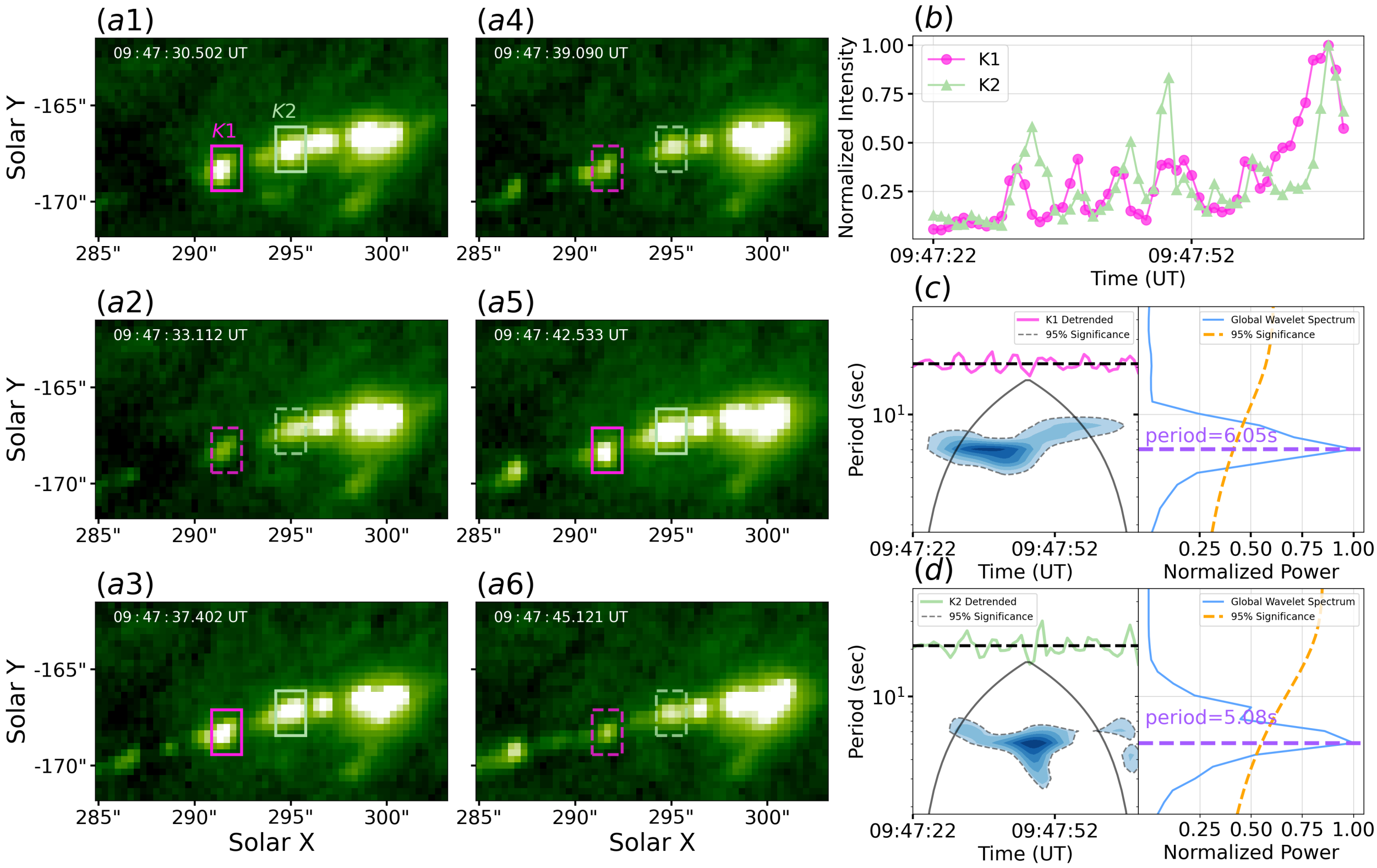

Fig. 2.— "Winking" ribbon kernels of Event 1. (a1)-(a6) *IRIS* SJI 1330 Å images showing the intensity oscillation of two ribbon kernels "K1" and "K2". Solid and dashed rectangles show the "bright" and "dark" times, respectively. (b) The integrated intensity evolution for two ribbon kernels "K1" and "K2". The regions used for calculating the intensity profiles are denoted by two rectangles in panels (a1)-(a6). (c)-(d) The wavelet analysis of integrated intensity profiles for "K1" and "K2", including the wavelet power map of the detrended intensity and corresponding global power. The orange dashed lines represent the 95% confidence level, and the purple dashed lines denote the period over the 95% confidence level with the maximal power. An animation showing the zoomed flare ribbon from 09:46 to 09:50 UT is available. The real-time duration of the animation is 24 s.

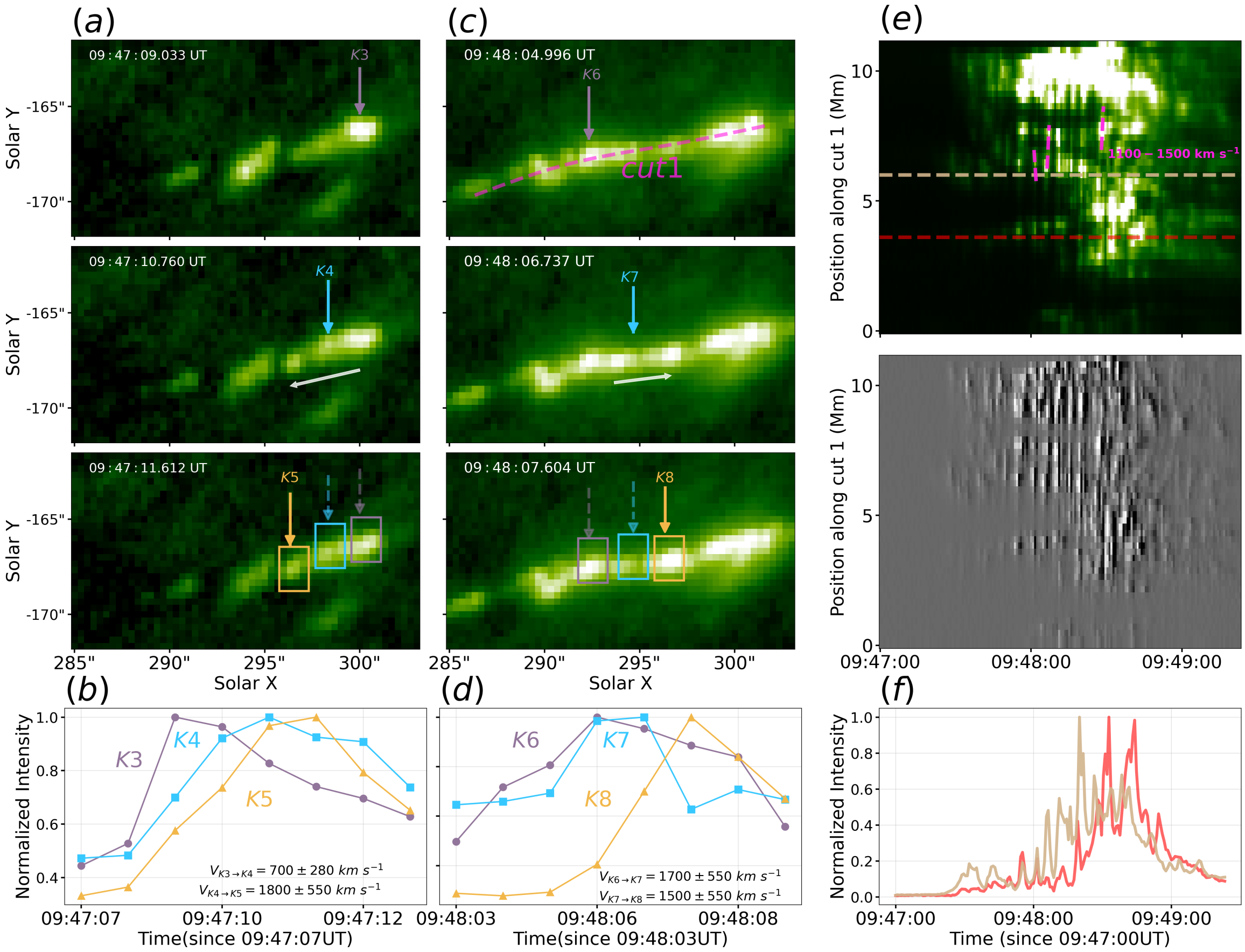

Fig. 3.— Fast slippage of ribbon kernels for Event 1. (a) IRIS SJI 1330 Å images showing one episode of the apparent slipping motion from kernel "K3" to "K5". The solid colored arrows mark the approximate location of the kernel along the direction of its motion (white arrow), while the dashed colored ones indicate its approximate location in the previous SJI snapshot. (b) The integrated intensity profiles for kernels "K3" – "K5" calculated within the colored rectangles in panel (a). (c) IRIS SJI 1330 Å images showing the slipping episode from kernel "K6" to "K8". (d) The integrated intensity profiles for kernels "K6" – "K8" calculated within the colored rectangles in panel (c). (e) Stack plot along cut 1 (magenta dashed line in panel (c)) and the corresponding running-difference stack plot. (f) Horizontal slices along two horizontal dashed lines in panel (e).

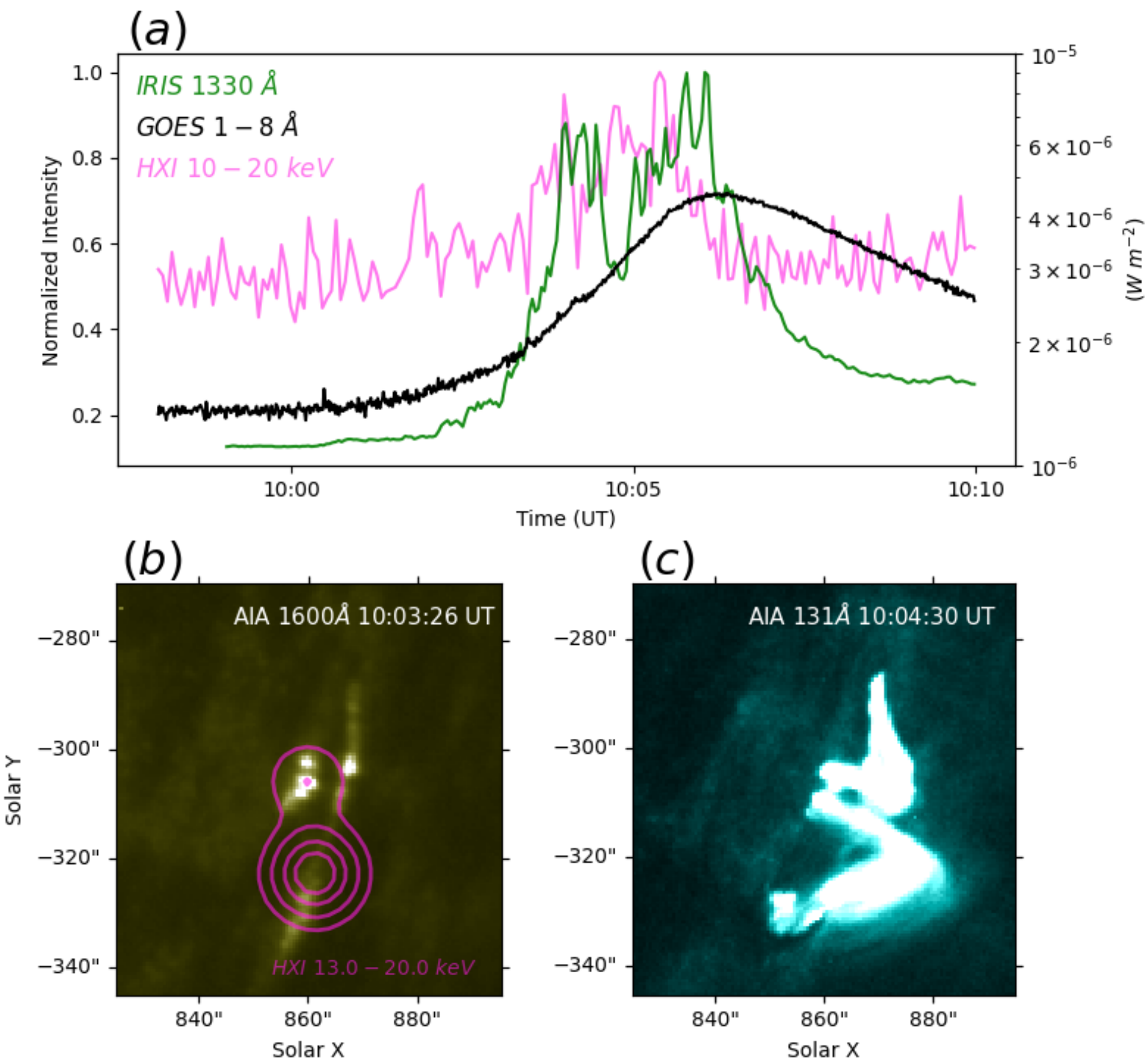

Fig. 4.— Overview of Event 2 on 2024 September 23. (a) Normalized HXR count fluxes of 10−20 keV range observed by HXI (red curve), IRIS SJI 1330 Å integrated intensity (green curve) within FOV of Figure 5(a) and soft X-ray flux in the 1−8 Å channel of the GOES satellite (black curve). (b) AIA 1600 Å image showing two flare ribbons. HXR contours of 13−20 keV band are overlaid. The contour levels are 10%, 25%, 50%, and 75% of the peak intensity. (c) AIA 131 Å image showing high-temperature flare loops connecting two ribbons.

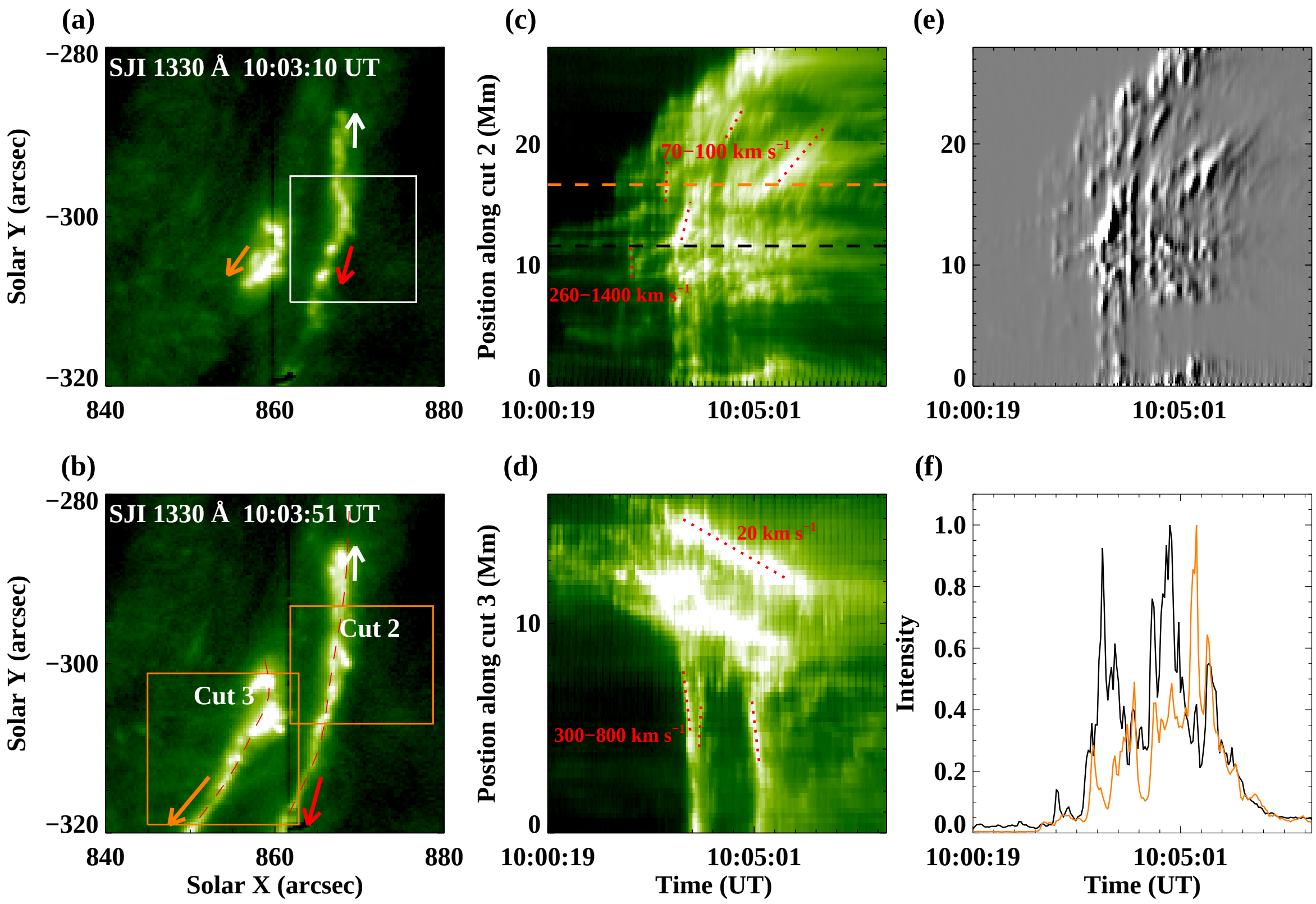

Fig. 5.— Slipping motions of ribbon kernels of Event 2. (a-b) IRIS SJI 1330 Å images showing apparent slipping motions of two flare ribbons. (c-d) Two stack plots along cuts 2 and 3 in panel (b), respectively showing the slipping motions along the western and eastern ribbons. The pulsations at the upper part of the stack plot in panel (d) are caused by the scanning slit and are ignored in this work. (e) Running-difference stack plot along cut 2. (f) Horizontal slices along two horizontal dashed lines in panel (c). An animation showing the evolution of the flare ribbon from 10:01 to 10:07 UT is available. The real-time duration of the animation is 14 s.

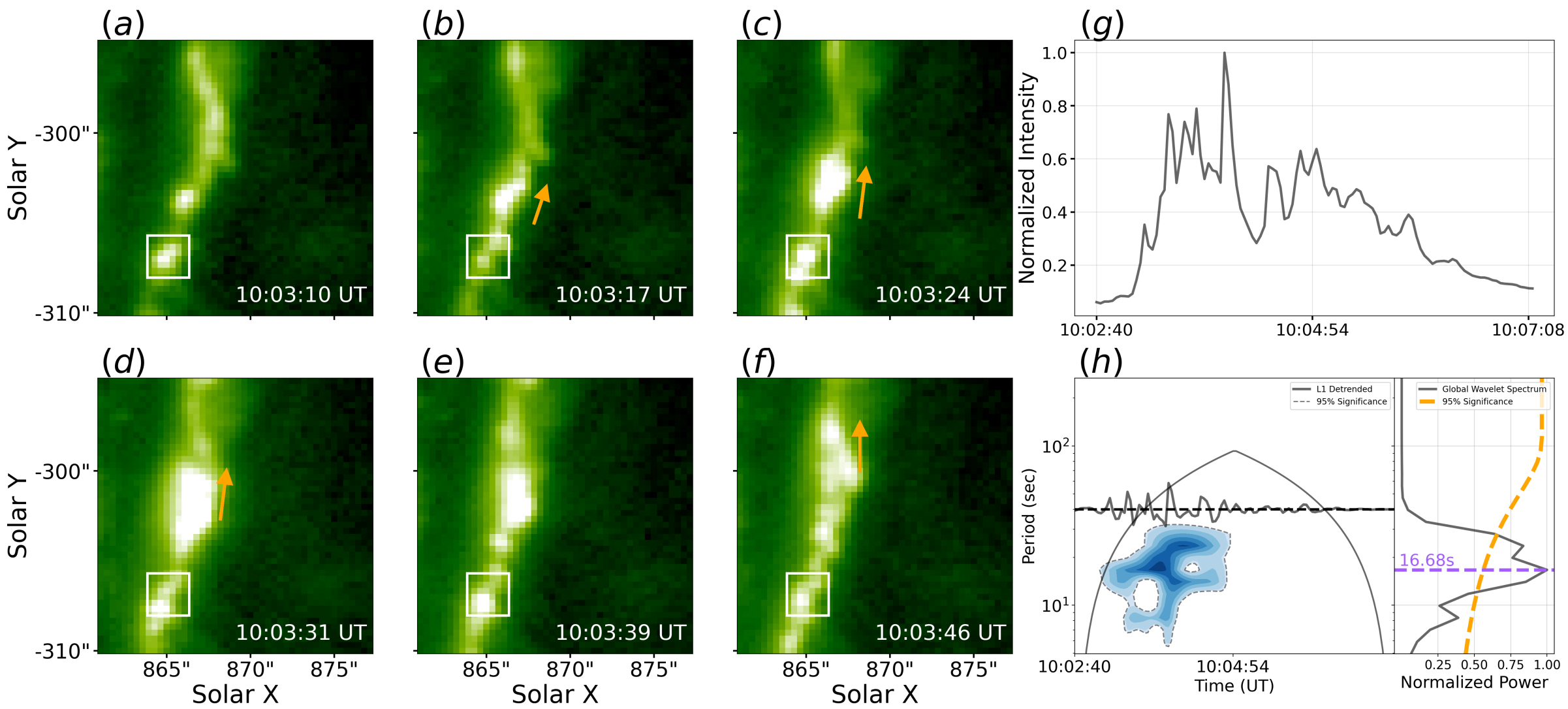

Fig. 6.— "Winking" process of the kernels in Event 2. (a)-(f) IRIS SJI 1330 Å images simultaneously showing the "winking" and slipping processes of the western ribbon. The FOV is shown by the white square in Figure 5(a). Panels (a), (c) and (e) show the "bright" times at white squares. Orange arrows denote the apparent slipping motion of other kernels. (g) The integrated intensity evolution for the kernels in white squares in panel (a)-(f). (h) The wavelet analysis of integrated intensity profile in panel (g), including the wavelet power map of the detrended intensity and corresponding global power. The orange dashed lines represent the 95% confidence level, and the purple dashed lines denote the period over the 95% confidence level with the maximal power.

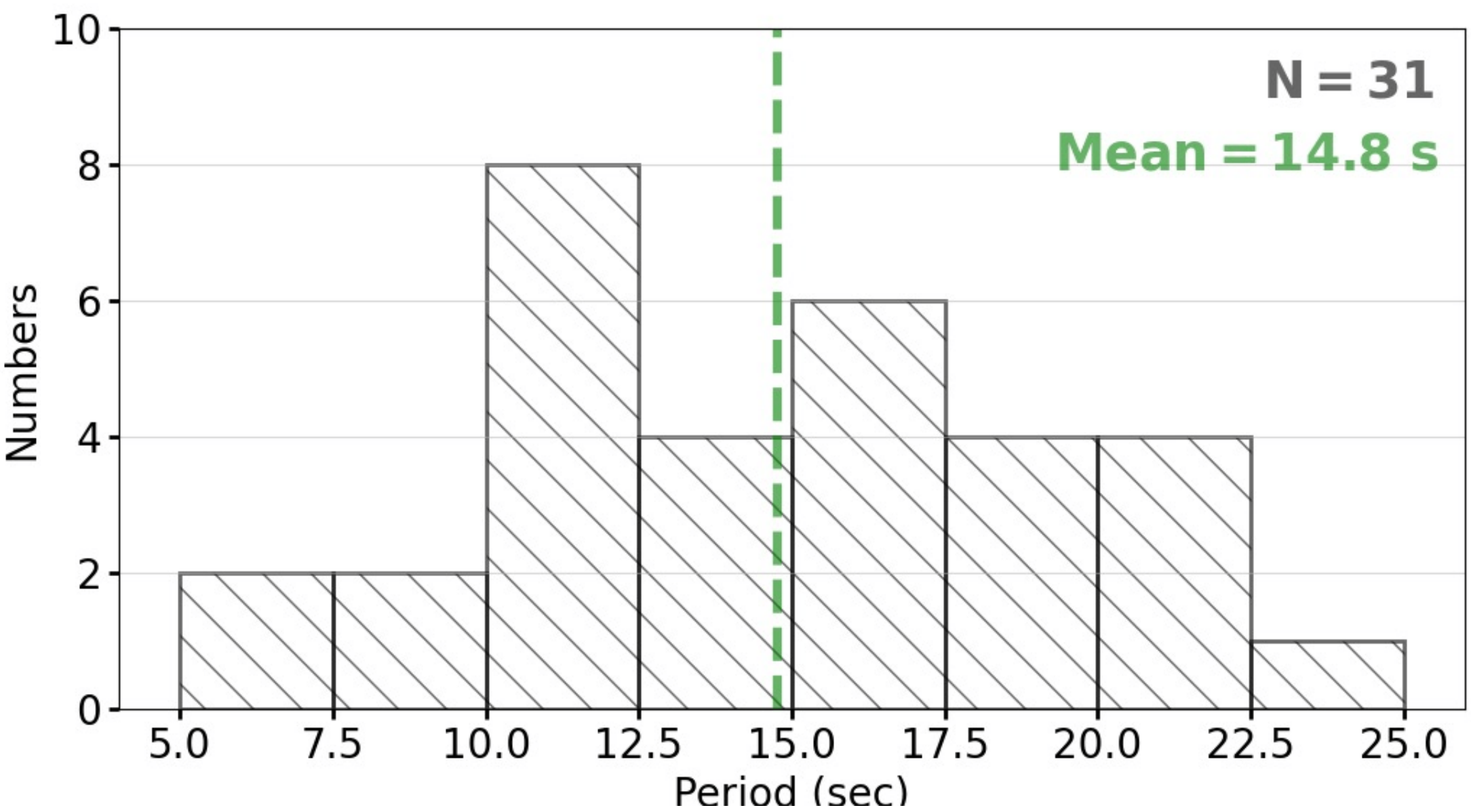

Fig. 7.— Histogram of identified QPP periods from the statistical 31 events including 28 C-class, 2 M-class and 1 unrecorded solar flares. The period range is 6−24 s and the mean period is 14.8 s (vertical green line).

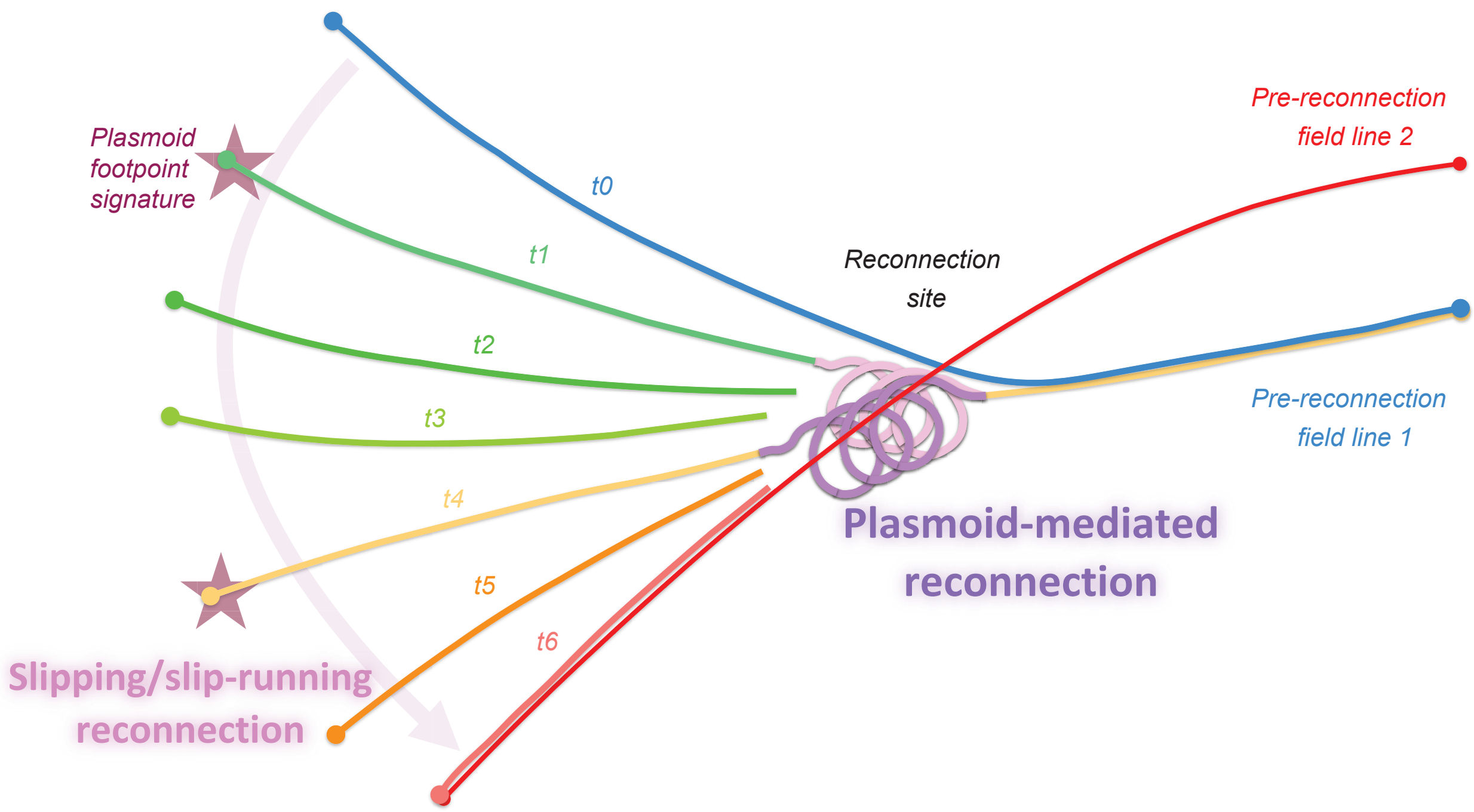

Fig. 8.— Cartoon depicting the combination of slipping/slip-running and plasmoid-mediated reconnection. Magnetic field lines (colored lines) undergo a continuous series of slipping/slip-running and plasmoid-mediated (bursty) reconnections and display an apparent slipping motion from "t0" to "t6". The slippage direction is indicated by the pink arrow. The twisted field lines show the plasmoids at the reconnection current sheet. The plasmoid footpoint signatures are denoted by colored stars.

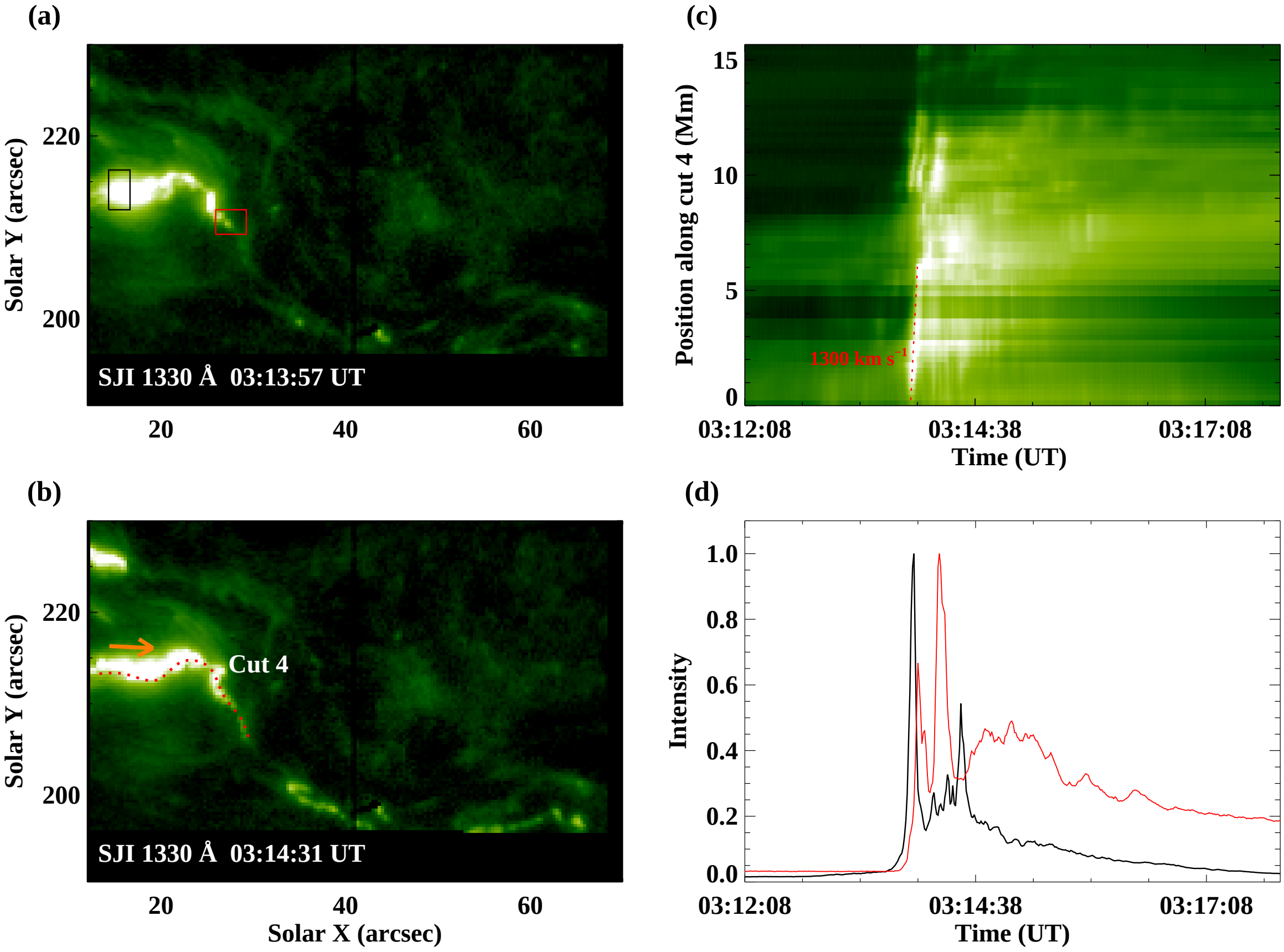

Fig. 9.— QPPs and slipping motion of Event 3 on 2025 February 05. (a)-(b) *IRIS* SJI 1330 Å images showing two flare ribbons. Orange arrow denotes the slipping direction of ribbon kernels. (c) Stack plot along cut 4 in panel (b) showing multiple-stripe pattern. (d) The integrated intensity profiles within black and red rectangles in panel (a). The period of the pulsations is about 9 s. An animation showing the evolution of the flare ribbon from 10:01 to 10:07 UT is available. The real-time duration of the animation is 14 s.

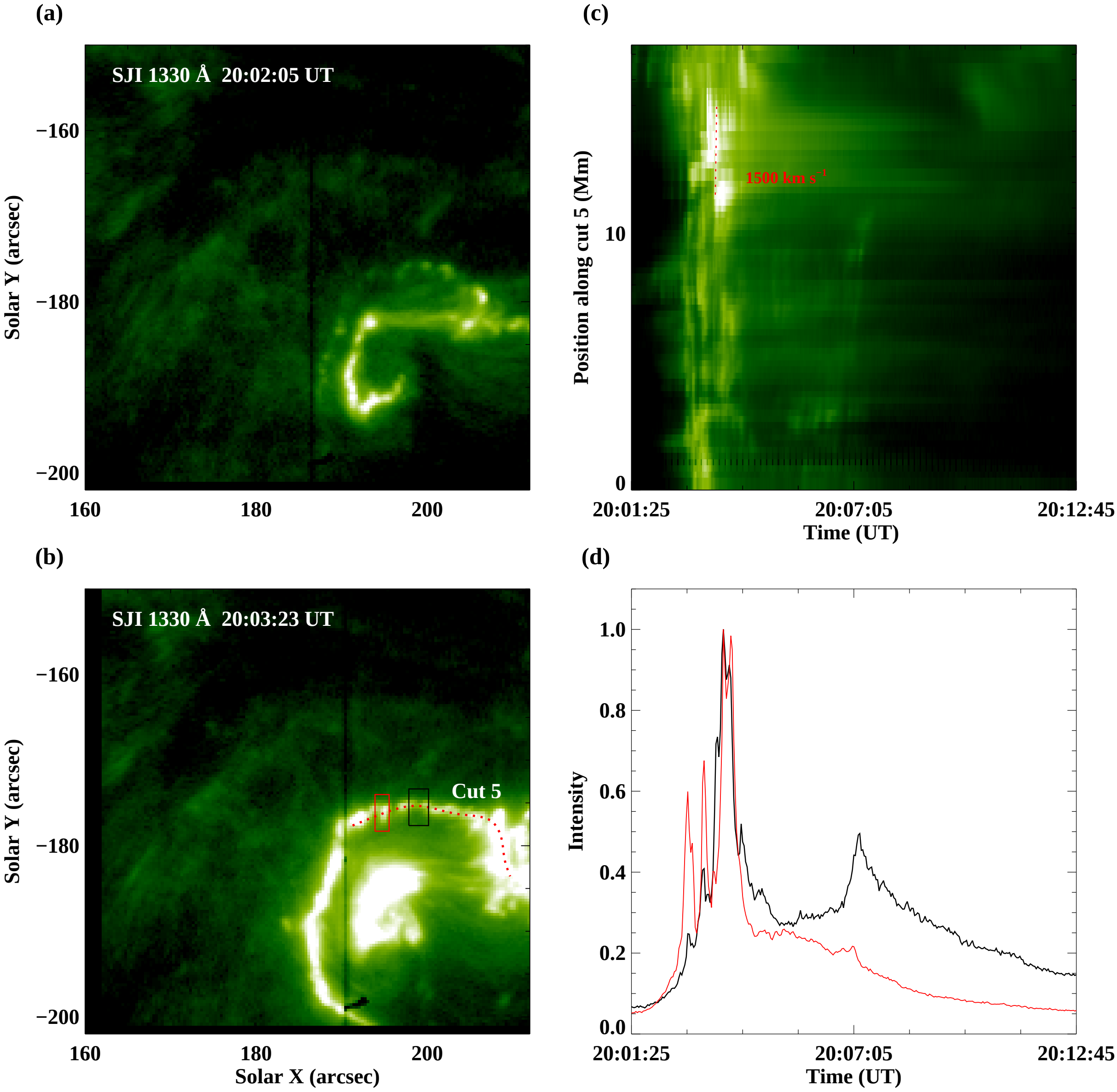

Fig. 10.— QPPs and slipping motion of Event 4 on 2025 April 05. (a)-(b) *IRIS* SJI 1330 Å images showing the circular flare ribbon. (c) Stack plot along cut 5 in panel (b) showing multiple-stripe pattern. (d) The integrated intensity profiles within the black and red rectangles in panel (b). The period of the pulsations is about 19 s. An animation showing the evolution of the flare ribbon from 03:13 to 03:15 UT is available. The real-time duration of the animation is 16 s.

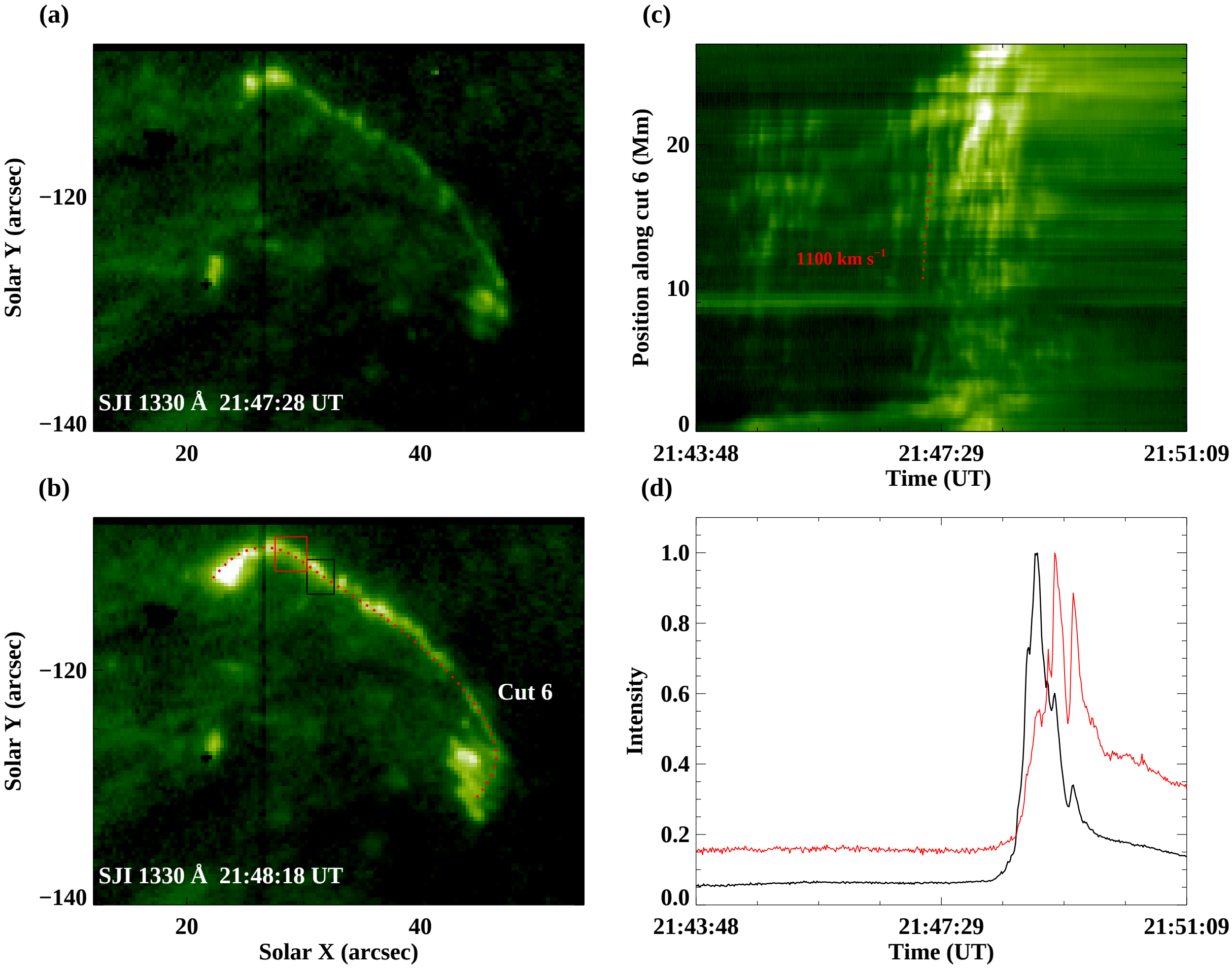

Fig. 11.— QPPs and slipping motion of Event 5 on 2024 November 23. (a)-(b) *IRIS* SJI 1330 Å images showing the circular the flare ribbon. (c) Stack plot along cut 6 in panel (b) showing multiple-stripe pattern. (d) The integrated intensity profiles within the black rectangle and red square in panel (b). The period of the pulsations is about 11 s. An animation showing the evolution of the flare ribbon from 21:45 to 21:50 UT is available. The real-time duration of the animation is 32 s.

Table 1: List of 31 flare events in our dataset

| INDEX | Date | Time [UT] | Cadence [s] | Solar X [arcsec] | Solar Y [arcsec] | ARNUM[1] | Flare class[2] | Period [s] |
|---|---|---|---|---|---|---|---|---|
| 1 | 20161012 | 11:50-11:59 | 1.68 | 582 | -322 | 12599 | C1.0 | 12 |
| 2 | 20201203 | 09:55-10:05 | 1.68 | -648 | -404 | 12790 | N | 14 |
| 3 | 20220406 | 03:49-03:58 | 1.70 | 646 | -211 | 12978 | C1.1 | 17 |
| 4 | 20220925 | 06:09-06:36 | 1.86 | -330 | -486 | 13107 | C4.3 | 18 |
| 5 | 20220925 | 07:01-07:10 | 1.88 | -330 | -486 | 13107 | C3.8 | 18 |
| 6 | 20230118 | 15:24-15:38 | 1.71 | -255 | 368 | 13192 | C3.4 | 12 |
| 7 | 20230209 | 05:00-05:20 | 1.86 | 178 | 582 | 13213 | C6.0 | 13 |
| 8 | 20230505 | 15:20-16:31 | 1.89 | -423 | 293 | 13296 | C7.6 | 22 |
| 9 | 20230520 | 18:36-18:58 | 1.88 | -682 | 305 | 13311 | C7.1 | 22 |
| 10 | 20231105 | 20:44-20:52 | 0.77 | -674 | -220 | N | C1.2 | 11 |
| 11 | 20231123 | 19:11-19:27 | 0.97 | -409 | 311 | 13492 | C3.8 | 11 |
| 12 | 20231129 | 15:11-15:35 | 0.85 | 138 | -331 | 13500 | C3.0 | 7 |
| 13 | 20240209 | 10:16-10:32 | 2.00 | -197 | -139 | 13576 | C6.2 | 16 |
| 14 | 20240825 | 09:45-09:51 | 0.87 | 256 | -161 | 13796 | C3.0 | 6 |
| 15 | 20240921 | 10:50-10:59 | 2.30 | 632 | -336 | 13825 | C1.2 | 22 |
| 16 | 20240923 | 09:59-10:18 | 2.40 | 849 | -295 | 13825 | C4.5 | 17 |
| 17 | 20240924 | 09:21-09:41 | 1.12 | -618 | -439 | N | C3.2 | 11 |
| 18 | 20241123 | 21:43-21:57 | 0.97 | 23 | -139 | 13901 | C3.8 | 11 |
| 19 | 20250117 | 07:45-08:01 | 2.20 | -639 | -102 | N | C4.6 | 13 |
| 20 | 20250130 | 02:04-02:17 | 0.97 | -759 | 266 | 13976 | C2.3 | 10 |
| 21 | 20250204 | 23:55-00:10 | 0.88 | 10 | 224 | 13981 | C5.2 | 17 |
| 22 | 20250205 | 03:07-03:28 | 0.87 | 40 | 225 | 13981 | M1.1 | 9 |
| 23 | 20250301 | 10:50-11:00 | 0.90 | 187 | 416 | 14006 | C2.5 | 14 |
| 24 | 20250317 | 15:49-16:11 | 0.87 | 789 | 140 | N | C6.5 | 9 |
| 25 | 20250405 | 19:50-20:25 | 2.27 | 142 | -171 | 14048 | M1.0 | 18 |
| 26 | 20250725 | 07:14-07:24 | 2.28 | -7 | 185 | 14149 | C1.2 | 18 |
| 27 | 20250730 | 06:47-07:11 | 2.40 | 185 | -312 | N | C1.6 | 20 |
| 28 | 20250831 | 01:20-01:49 | 2.40 | 216 | -371 | N | C5.6 | 17 |
| 29 | 20251011 | 19:10-19:25 | 2.28 | 534 | -258 | 14247 | C2.6 | 24 |
| 30 | 20251025 | 10:16-10:38 | 2.40 | -714 | 11 | 14267 | C1.5 | 12 |
| 31 | 20251109 | 00:27-01:27 | 2.43 | -95 | 351 | N | C1.7 | 17 |

[1] NOAA Active Region Number, N means no AR recorded

[2] Flare class based on the peak soft X-Ray flux, N means no flare recorded